# Modelling of Networked Measuring Systems
## - From White-Box Models to Data Based Approaches -

Klaus-Dieter Sommer[1], Peter Harris[2], Sascha Eichstädt[3], Roland Füssl[1], Tanja Dorst[4], Andreas Schütze[4], Michael Heizmann[5], Nadine Schiering[6], Andreas Maier[7], Yuhui Luo[2], Christos Tachtatzis[8], Ivan Andonovic[8] and Gordon Gourlay[9]

[1] Technische Universitaet Ilmenau, Germany
[2] National Physical Laboratory, Teddington, United Kingdom
[3] Physikalisch-Technische Bundesanstalt, Braunschweig and Berlin, Germany
[4] ZeMA - Zentrum für Mechatronik und Automatisierungstechnik gGmbH, Saarbrücken, Germany
[5] Karlsruhe Institute of Technology, Germany
[6] Centre for Measurement and Calibration, Wolfen, Germany
[7] Friedrich-Alexander-University of Erlangen-Nuremberg, Germany
[8] Department of Electronic and Electrical Engineering, University of Strathclyde, UK
[9] Advanced Forming Research Centre, University of Strathclyde, UK

*Correspondence to*: Roland Füßl (roland.fuessl@tu-ilmenau.de)

**Abstract.** Mathematical modelling is at the core of metrology as it transforms raw measured data into useful measurement results. A model captures the relationship between the measurand and all relevant quantities on which the measurand depends, and is used to design measuring systems, analyse measured data, make inferences and predictions, and is the basis for evaluating measurement uncertainties. Traditional modelling approaches are typically analytical, for example, based on principles of physics. But with the increasing use of digital technologies, large sensor networks and powerful computing hardware, these traditional approaches are being replaced more and more by data-driven methods. This paradigm shift holds true in particular for the digital future of measurement in all spheres of our lives and the environment, where data provided by large and complex interconnected systems of sensors are to be analysed. Additionally, there is a requirement for existing guidelines and standards in metrology to take the paradigm shift into account. In this paper we lay the foundation for the development from traditional to data-driven modelling approaches. We identify key aspects from traditional modelling approaches and discuss their transformation to data-driven modelling.

# 1 Introduction

Economies and society are undergoing a significant development regarding the use of technology in the form of a so-called *digital transformation*. Especially in industry, this development is changing production processes at an increasing pace. In the vision of the *Internet of Things* (IoT)*,* the transformation is realised as fully digitised processes in which complex networks of sensors, which can be adjusted to particular tasks, are used for monitoring, control and prediction. The networks often include an internet-based connectivity of the sensors, similar to the *Industrial Internet of Things* (IIoT). Due to the increasing availability of low-cost IIoT-enabled measuring devices, sensor networks are much larger than is usually the case for traditional measurement applications. As a consequence of the size and complexity of the networks, data analysis becomes more and more data-driven and, in particular, machine learning methods are often applied. Moreover, the use of *digital twins*, in which a model of a physical object is updated based on an evolving set of data [Wright 2020], and software sensors as a fusion of models with sensor data, increases the role of modelling in the Factory of the Future. With this paradigm shift in the treatment and analysis of measured data, traditional approaches for modelling in metrology need to be reconsidered. In addition, for the effective control of production and delivery in supply chains, such as in-process control, for quality control and for economic purposes, process modelling is essential. Figure 1 shows the fields of application for modelling tasks with the example of the Factory of the Future and especially the role of modelling to support those fields.

*Mathematical modelling*, in general, involves the assignment of mathematical terms for all the relevant components of a system or process and the derivation of mathematical equations giving the relationships between those terms. When applied in the context of measurement, one distinguishes between terms that relate to quantities that are known or measured and those that are unknown and to be estimated from measured data. The former can include quantities representing the responses of individual sensors, quantities representing the calibrations of those sensors as well as applied corrections to account for environmental effects. The latter can include quantities representing the stimuli to the sensors or representing system characteristics derived from those stimuli, and will include the *measurand*, the quantity intended to be measured [VIM 2012]. A model can be constructed using physical principles based on a theory that defines how the quantities depend on each other. Alternatively, a data-driven model is one in which a relationship between the quantities is expected or observed from data, but without a supporting theory. Many models have both physical and data-driven components. These are also known as hybrid models.

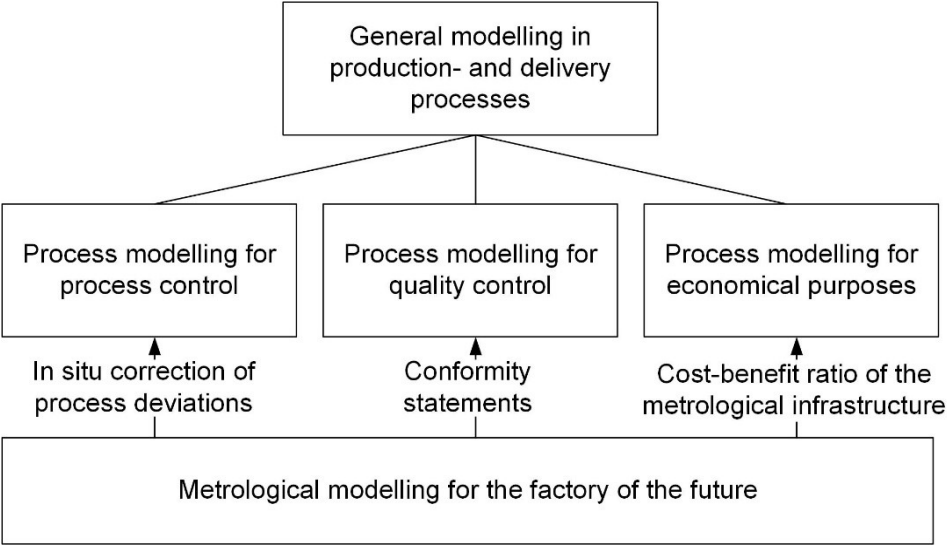

**Figure 1: Modelling tasks using the example of the Factory of the Future.**

Mathematical modelling is a key activity in support of metrology and applications dependent on measurement. It is used to inform the design of sensors and measuring systems, to make inferences and predictions about quantities of interest, and to increase our understanding of real-world systems or processes for which a model provides an abstract representation, albeit one that is necessarily approximate. The recently published "Guide to the expression of uncertainty in measurement — Part 6: Developing and using measurement models" [GUM6 2020] provides an overview of typical modelling approaches in metrology and their use for the evaluation of measurement uncertainty.

Models are useful to create an understanding of the mode of operation and the structure of the system under consideration. This understanding can then be used to generalize the mode of operation of the system to other scenarios. Models thus also form the basis of *simulation*. In system theory, which also includes control engineering, models are the basis for the description, analysis and synthesis of dynamic system behaviour. Critically, models provide the basis for evaluating *uncertainty*, which is required to understand the quality of quantity estimates, to use those estimates for decision-making, and to establish traceability of measurement results to primary standards. In metrology, the "Guide to the expression of uncertainty in measurement" (GUM) [GUM 2008] is considered to be a *de facto* standard and provides a framework for evaluating measurement uncertainty that is both model-based and probabilistic.

The current trend towards the digital transformation of manufacturing industry is driven by the deployment of large networks of so-called "smart" sensors and artificial intelligence to automatically make decisions about, and manage, production processes. This transformation presents a number of modelling challenges. Notably, the absence of physical models is compensated by the availability of large volumes of sensor data leading to a dependence on data-driven models, and particularly those implementing methods of machine learning and deep learning. However, a data-driven approach still needs explainable and traceable data streams and uncertainty evaluation to ensure confidence in the measurement results obtained. Furthermore, in a data-driven approach it is also necessary to address additional aspects such as sensor redundancy [Kok 2020a, Kok 2020b], timing and synchronisation issues [Jagan 2020], mixed-quality sensors, etc. associated with sensor networks.

In this paper we review the application of traditional, analytic-parametric modelling to dynamic and distributed measuring systems, with a focus on the accompanying well-established and documented methods for parameter estimation and uncertainty evaluation. We then motivate the application of data-driven modelling to such systems and discuss how to apply such modelling in a way that ensures that those systems can be considered as "metrology systems" supported by the necessary concepts of traceability and uncertainty quantification.

The paper is organised as follows. In section 2, we review the general objectives and components of mathematical models and then, in section 3 we discuss the ways in which those objectives and components apply in the context of measurement tasks. Approaches to mathematical modelling, including analytic-parametric, data-driven and combined approaches, are described in sections 4, 5 and 6, respectively, with an emphasis on how the development of such approaches reflects the digital transformation of manufacturing industry. In section 7, we illustrate the application of the different approaches to two manufacturing problems, viz., condition monitoring and predictive quality. Finally, section 8 contains our conclusions and outlook.

## 2 General objectives and concepts of mathematical modelling

### 2.1 General objectives of modelling

Generally speaking, a model is an image of a reality – in our case mostly the image of a process or item or system – reduced to the essentials. A model can be used to determine a measurement result and, consequently, to explain or to quantitatively simulate or to evaluate a process or system. A model can also be used to design a new sensor and, consequently, to design and create a system or process flow in a systematic and explainable way. Explainability and quantitative assessment require a (sometimes very) simplified view of real systems and processes. All models are imperfect and their ability to represent real-world behaviour is always limited. However, modelling must consider all the essential components and influences that determine the desired result. This interplay between simplification on the one hand and relevance and fitness-for-purpose on the other is known as the *pragmatism principle*.

Usually a model consists of a set of mathematical equations involving at least two quantities. In general, a model can be presented and applied graphically, in tabular form, as a flow chart or schedule, as explanatory text or in other ways.

### 2.2 General sequence of modelling steps

In the "pure" modelling literature, modelling is divided into six steps, as follows:
a) **Delimitation:** Models are usually not universally applicable but refer to a specific scope. The working scope to which the model is applicable may be defined, e.g., by physical conditions, such as the linear range of a circuit, the laminar or turbulent velocity range of a flow or the temperature range in which a substance has a certain property. The scope of the model essentially determines its *generalizability*.

b) **Abstraction**: Models are usually not created for individual systems alone, but for classes of systems that can be described in the same way. A model can then be applied to a whole class of identical or similar systems – each after adjusting the model parameters. For example, the modelling of a pump by means of a characteristic curve can be applied to very different physical implementations of pumps provided that the tiling properties of the pumped substances belong to a certain class. Similarly, strong abstractions are found, e.g., in the description of electrical networks by resistances, inductances and capacitances.

c) **Reduction**: The step of abstraction is usually accompanied by the necessary reduction of the system description to make it understandable and explainable in terms of the objective. Influencing factors and details that are of little relevance for the system behaviour are omitted. Of course, this reduction of the system description and, thus, reduction of the information about the system inevitably results in deviations of the modelled behaviour from the real system behaviour to be evaluated. In the context of measurement technology, systematic or random deviations arising from the reduction are to be assessed, preferably by means of measurement uncertainty analysis.

d) **Decomposition**: It is generally the case that real, mostly complex, systems cannot be fully understood, consistently analysed, and adequately described mathematically, graphically or by other means. Then, the preferable approach is to decompose the system into manageable and easy to model sub-systems so that several sub-models represent in their entirety the overall system. The challenge here is to select the boundaries of the sub-systems in such a way that functional primitives are created that are as easy to model as possible, so-called *flat modelling*. The choice of the system boundaries of the sub-systems can be based on previous knowledge about the structure of the system and to what extent internal variables of the overall system are accessible as inputs and outputs of the sub-systems. Another approach to define sub-systems is to classify them according to the physical domains to which they apply.

e) ***Aggregation*:** To generate the model for the overall system, the sub-models generated in the decomposition step must be combined to form an overall model. This step requires that the sub-models complement each other in such a way that a valid overall model can be created. Excessive reduction within the sub-models can result in deviations in the overall model that are still considered tolerable for each sub-model. These deviations can have such a strong impact on the overall model that the model formation must be revised, e.g., by including influencing variables or nonlinearities that have not been considered so far. Aspects of the stability of dynamic models must also be considered in this step.

f) ***Verification*:** In this last step, the deviations between model and reality are to be empirically determined and evaluated. In addition to the evaluation of systematic and random deviations, questions of system dynamics, the temporal variability of system behaviour and the system reaction to external disturbances play a major role. As a matter of course, often the model is adapted successively to the desired capabilities, and decomposition into simple partial models (see above) also opens up the possibility of adapting the model to changing realities.

**2.3 Forward and backward modelling in measurement**

The classical cognitive-systematic approach to understanding an object, system or process leads from the cause(s) to the effect(s), or in the time domain from the beginning or starting time to the end of the temporal process [Sommer 2005, Sommer 2006]. This approach is usually the way explanations and justifications, teaching materials and much more are structured. The approach is also the usual and most common way of model building in the natural sciences and in technology, not least because of the analogous human thinking. An example is systems analysis, and it is also the classical way of representing measuring systems and signal flows, from the main *cause*, the quantity to be measured with its largely unknown value, to the *effect*, the displayed or signal output value of the measuring system (see Figure 2).

      In the context of metrology, the starting point of modelling is usually a measuring chain whose individual elements, obtained from the decomposition of the overall measurement, are understood as transmission elements. The measurand is the main input, the first transmission

elements are necessarily transducers or sensors, and the chain ends with the display or signal output. Influencing or disturbance variables can be seen as secondary input variables. Figure 2 illustrates the approach of *cause-and-effect modelling* with the example of a dimensional measuring process. The individual transmission elements can initially be considered separately from each other. This consideration includes both known physical relationships for the transmission behaviour, e.g., to consider optical surface properties of the test specimen, as well as empirical relationships, e.g., thermal expansion or the temperature dependence of the measuring object. Such a cause-and-effect model of a measurement could be used, e.g., to simulate the essential and significant behaviour of a sensor or a measuring system under development. In metrology, cause-and-effect modelling, or *forward modelling*, is often based upon fundamental physical relationships and corresponding material measures used to generate or reproduce reference values of quantities.

But by far the most common applications in metrology naturally aim at providing the (best estimated) value of the measurand and the evaluation of its assignable uncertainty that is a deduction from the displayed output value to the value of the measurand, which is the cause at the input of the model of the measuring chain. A backward or inverse model serves to infer information about the measurand from the indicated value and knowledge about any dominant influence quantities and the modelled process. Therefore, evaluating a measurement involves solving an *inverse problem*. The backward model is also called the measurement equation [VIM 2012] or evaluation model or measurement model [GUM 2008]. It is obtained by inverting the cause-and-effect or forward model and is the starting point for applying the methods of uncertainty evaluation described in the GUM [GUM 2008] and its supporting documents, e.g., [GUM1 2008]. The derivation of the measurement equation from the cause-and-effect relationship is illustrated in the lower part of Figure 2.

It is important to note that data-driven, especially adaptive and learning models, can usually be taught directly in the "reverse direction". Inversion of forward models is generally possible, although doing so may be numerically unstable, and in order to alleviate such instability regularization techniques allow more robust inversion schemes [Vogel 2002].

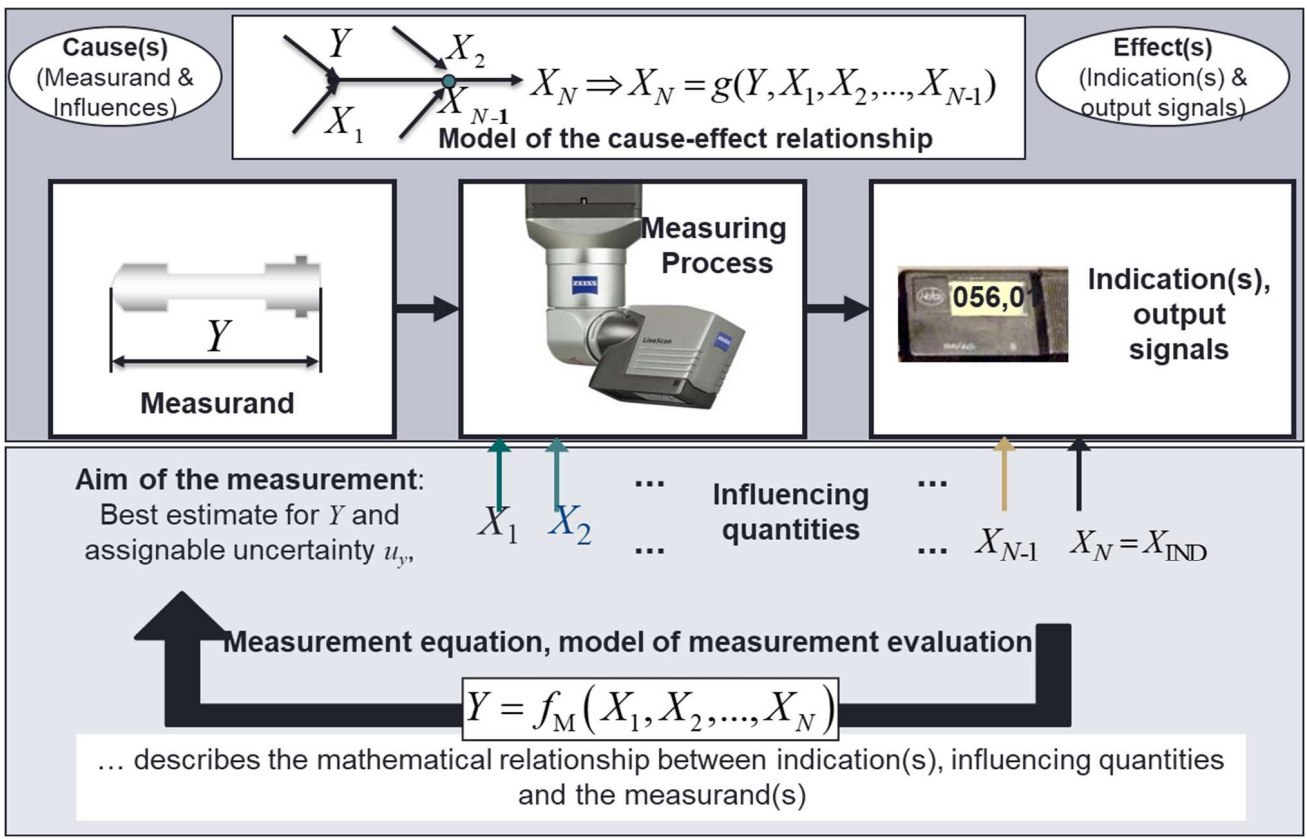

Figure 2: Modelling a measurement: (top) cause-and-effect relationship and (bottom) measurement equation [Sommer 2018].

## 2.4 White, grey and black box models

The mathematical terms used to represent the relevant components or quantities of a system or process will generally include *model parameters* that are used to capture the dependencies between those terms. The nature and meaning of such parameters vary with the type of model. For instance, in a model derived from physics, the model parameters typically refer to certain physical properties, such as a temperature coefficient, mass or time. Such models are referred to as *white box models*. Sometimes, the dependencies cannot be derived from physical arguments alone. For instance, a straight line fit through a point cloud may be derived from correlation arguments rather than from a pure physical reasoning. The parameters obtained in this way may still carry a physical interpretation, though. Hence, models of this type are a mixture of a direct relation based on a known (physical or other) relationship and a more abstract relation. Such models are referred to as *grey box models.* When there is no direct relation derived from physical or other arguments, the model is typically referred to as *black box model*. Figure 3 shows an overview of these three types of model.

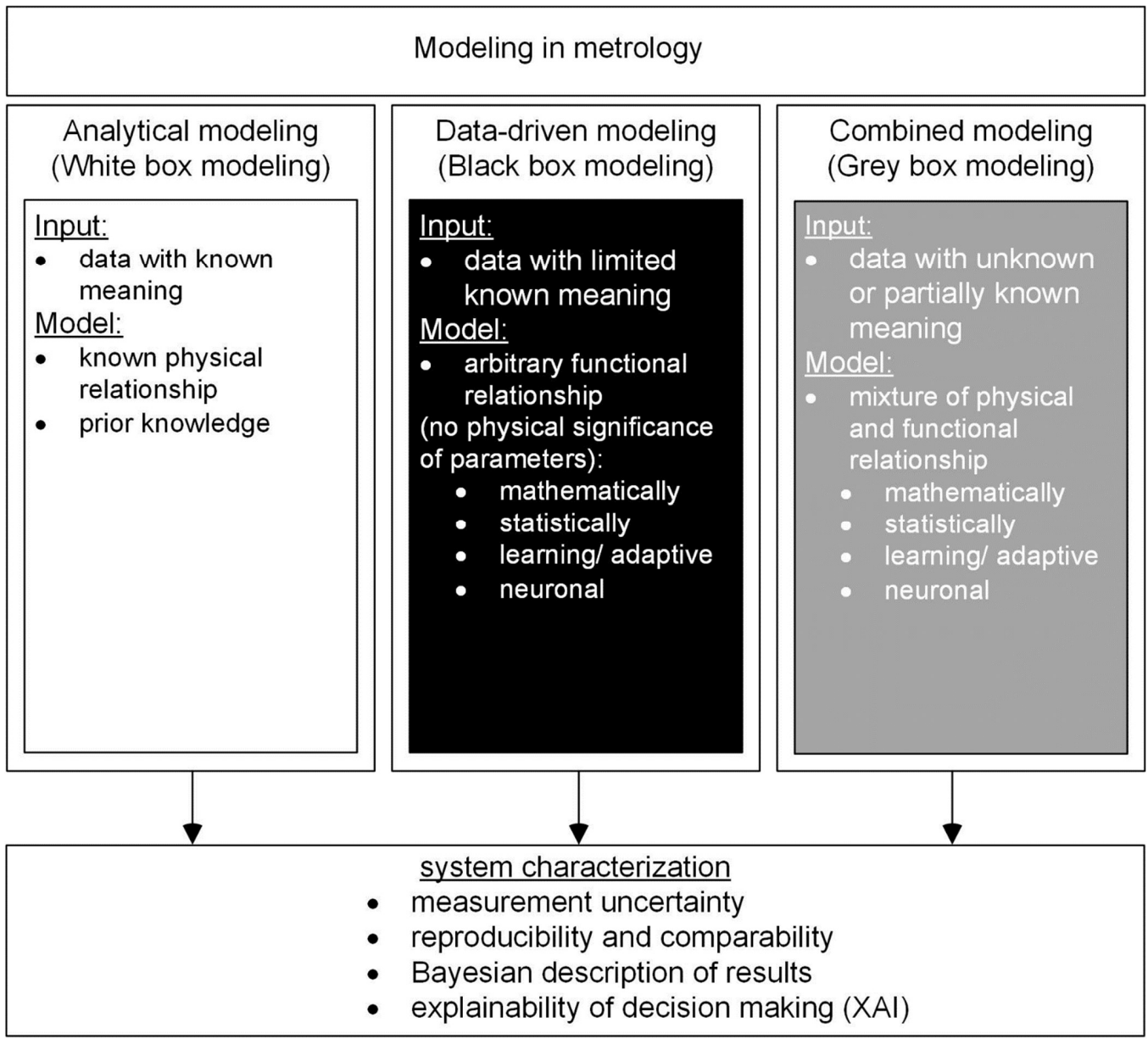

Figure 3: White, grey, and black box models and their basic properties.

The general principles of modelling, as described before, apply equally to white, grey and black box models. But depending on the choice of model, generalization and transferability may be challenging as learning theory generally only guarantees stability if the training data comes from the same distribution as the test data. In particular, parameter-rich black box models typically suffer from such problems (see section 5.1). In order to compensate for data mismatch, the discipline of *transfer learning* or *adaptation* was developed, which helps to support the generalization of black box models [Tan 2018, Kouw 2019]. However, domain transfer is still more difficult than for classical methods of modelling due to the limited interpretability of the learned parameters.

## 3 Model requirements in the Digital Future of Metrology

### 3.1 General measurement tasks in Future Metrology

In the context of the Future Metrology, different measurement tasks can arise. One example involves sensors that are deployed within a space, such as a test cell, laboratory, or workshop, and are used to estimate the measurand as a function of spatial position and possibly time. The sensors may differ in quality and performance, some may be calibrated and others uncalibrated, and the sensors may communicate with each other. This scenario arises, for instance, when reference temperature sensors are located at different positions in space and are used to establish a map of estimates of temperature as a function of spatial position. This map can in turn be used for the calibration of other temperature sensors that are not necessarily co-located with the reference sensors. For such a calibration task, however, the temperature map is insufficient, but must be accompanied by information about the uncertainty of an estimate of temperature at any spatial position and the covariance of estimates at any two different spatial positions. Established documents on measurement uncertainty, such as the *GUM*, do not provide guidance on how to treat such a problem in which the measurand is a function.

Another example involves estimating a measurand such as a single property of a manufactured workpiece or manufacturing process, or a measurand defined at a single point in space and time, or a measurand comprising finitely-many such quantities, using sensors that are distributed in space and measuring at different times. Consider, e.g., sensor measurements for a production process that are used to predict the geometric dimensions of the resulting workpiece and, subsequently, used to optimise the production process. Another example is the lifetime assessment of a mechanical component, where measurements of various physical quantities are analysed in order to estimate the remaining lifetime of the component.

Yet another example is to use sensor measurements to estimate a measurand whose possible values are labels that define membership of different classes. Consider, e.g., the task of using those measurements to decide whether a manufactured workpiece is in or out of tolerance based on a certain definition of quality classes, or whether a component will fail or not, or whether a workpiece surface contains anomalies based on whether they are present or not rather than quantified dimensionally. In contrast to classical approaches to using tolerances to make decisions, such as based on the statistical significance of the deviation of an uncertain measured value from a fixed nominal value, quality classes in the Factory of the Future may be purely data-driven, e.g., based on clustering of training data with known quality. The output of many data-driven models, such as machine learning and deep learning models, is a set of values that are often interpreted as describing the "likelihoods" of membership of each class. However, as above, established metrology documents do not give guidance on the treatment of such models, and the evaluation of the uncertainty associated with those likelihoods is a topic of current research. Moreover, the applicability of the above described modelling steps to such data-driven approaches needs to be assessed.

## 3.2 Measuring system characteristics in the Digital Future of Metrology

The provision of uncertainty information is necessary to establish the *traceability* of a measurement result, which involves relating the result to a reference such as a primary standard through an unbroken chain of calibrations with each link in the chain contributing to the measurement uncertainty [VIM 2012]. In this regard, metrological traceability is essential for the digital future of metrology and any every measurement. Measurement results only becomes useful and valuable when they are traceable and quality assured. Measurement traceability is important, for example, in the Factory of the Future, e.g., to ensure comparability between plants after changes in the underlying measurement infrastructure (e.g., replacing a faulty sensor). Traceability can be achieved only by a proper evaluation of measurement uncertainty and considering all relevant influence quantities in the modelling that underpins the evaluation.

Another important characteristic of measurements made in the context of the Metrology of the Future is that the quantities involved, notably the sensor responses and sensor stimuli, are generally *dynamic* in the sense that their values depend on time [Esward 2019]. Here, the time-dependence can be considered as continuous or as discrete and obtained by sampling continuous functions of time. Furthermore, the measured values of a dynamic quantity for different times are generally not independent, resulting in non-zero covariances that have to be considered when calculating estimates and evaluating their uncertainties. The class of linear time invariant (LTI) systems described in section 4 is the simplest class of model for dynamic quantities and is appropriate for a wide range of applications.

The need to consider dynamic quantities and, in particular, to treat *digital* information obtained by sampling continuous functions of time means that models arising in the context of the Factory of the Future must consider timing and synchronisation effects. Sensors may have access to an accurate clock but for multiple, spatially distributed sensors, it is essential that the sensors are working to a common timescale, e.g., through traceability to UTC, so that the data they record can be analysed correctly. Accurate timestamping can be an important diagnostic aid to establishing, e.g., which cause-effect relationships are feasible, and which are infeasible, or providing information about the path of a signal (such as vibrations) traversing a structure. In addition to the requirement to ensure that the data recorded by different sensors is synchronised, such data can be subject to timing errors, such as jitter. Pre-processing of the sensor signals can be applied to remove the effects of noise and jitter and correct for such timing errors. However, it is important to understand the uncertainty of the resulting pre-processed signals, and to be able to propagate that uncertainty through any subsequent processing or aggregation of the sensor signals.

This requirement holds true for other kinds of data pre-processing and is the prerequisite for the evaluation of uncertainty in any subsequent data processing steps. As many data pre-processing methods are related to signal processing, developments from the area of dynamic measurement analysis can be employed. In the literature, methods are available for the evaluation of uncertainty, for instance, in the application of the discrete Fourier and wavelet transforms, for digital filtering, piecewise linear approximation and interpolation.

The deployment of potentially many sensors to collect data, sometimes without careful consideration of the design of the measuring system, means that identification and removal of redundancies in the input data streams and the application of techniques of sensor fusion and dimensionality reduction become particularly important in the context of the Factory of the Future. Models are required to capture a "pipeline" from the sensor measurements to the measurand that includes the steps of dimensionality reduction, feature extraction and machine learning, and methods are needed to propagate the uncertainties of the sensor data through that pipeline.

The Digital Future of Metrology provides novel possibilities to incorporate such models for measurements and measuring devices. For instance, a mathematical model for measurements made with a particular measuring instrument can be translated into a *digital twin* of the instrument. Provided the digital twin uses suitable programming interfaces, the mathematical model can then be applied mostly automatically and updated continuously as data becomes available. Moreover, the digital twin can be employed as a so-called "soft sensor" to enable advanced fault detection and monitoring. Similarly, the whole data analysis, and thus use of models, in the Factory of the Future is mostly automated, because for the large sensor networks used in such environments, manual modelling and data analysis are infeasible. Thus, modelling for the Factory of the Future should be modular, i.e., instead of modelling the complete sensor network, smaller and more versatile model building blocks are defined and combined depending on the purpose of measurement.

# 4 Analytical-parametric modelling (white box models)

## 4.1 Basics

The basis for analytical modelling is the analysis of the system structure. This type of modelling is therefore also referred to as structural modelling. It attempts to represent the physical system as directly as possible as a mathematical model in terms of all relevant influencing factors and interactions. The result is a model whose properties are completely known. Such a model is therefore also called a white box model. The quantities considered for modelling are usually physical quantities such as pressures, temperatures, flow rates, irradiances, or angles.

The creation of an analytical model is only possible if a physical understanding of the system and all influencing factors is available to a sufficient degree. However, any influences that are not understood or cannot be observed or are too complex to model or observe can be described by random variables in the sense of a model reduction. The aspects of decomposition and aggregation are implicit in this type of modelling through the selection and combination of the sub-systems used.

Such an approach is appropriate if the system components with their properties can be explicitly specified for a new system to be synthesized. The selected system components with their respective system behaviour and the type of combination, e.g., in the form of a block diagram, then directly determine the basic system behaviour. Finite-element models, in which the interaction of the elements is described by certain physical relationships, also belong to the class of analytical models.

Even if the basic, e.g., dynamic, behaviour of the system is determined by the specification of the system structure, an analytical model can be adapted to physical reality by changing the model parameters, e.g., including amplifications, time constants, etc. The identification of the model parameters can be done by specific experiments, e.g., by applying impulse or step functions or harmonic oscillations, or during operation of the system by comparing the expected and the actual system response. If the desired quality of the system modelling is not achieved in this step of parameter adjustment, structural changes can be made to the model in order to include the causes of the deviations that have not yet been considered.

As prior knowledge for the creation of analytical models, knowledge of the system structure is required. Here, the level of detail of the known system structure determines the achievable accuracy of the model. Previous knowledge of the model parameters can also be included, whereby an empirical optimization of the model parameters, in the sense of a data-driven optimization, usually increases the quality of the model fit.

The advantages of analytical modelling are, first of all, the interpretability of the model by humans since the quantities and functional blocks usually allow a concrete physical interpretation. Such interpretation also often allows a simple check of the basic model properties for plausibility. By modelling concrete physical system properties, analytical models often enable a good system description over a large interval of values of the modelled quantities whereby, of course, the range of validity defined in the delimitation step during the model creation must be considered.

A disadvantage of analytical models is the required physical understanding of the system with regard to its structure, its function, and possibly existing subsystems, which usually requires expert knowledge of the system. If the system structure is not known from the outset, considerable effort may be required for system analysis and for parameterization of the model to determine the relevant influencing factors and cause-and-effect relationships or to carry out appropriate experiments.

## 4.2 Parametric modelling procedure

For the simple case of a measurement task involving a single sensor, we start with a forward model of the form

$$Z = \phi(Y, \boldsymbol{\beta}) \tag{1}$$

that describes how the sensor output $Z$ depends on the sensor stimulus $Y$ and sensor parameters $\boldsymbol{\beta}$. The parameters $\boldsymbol{\beta}$ might describe the gain and offset of the sensor or more general calibration parameters for the sensor.

If the sensor is calibrated, the measurement task can be to infer $Y$ from knowledge of $Z$ and $\boldsymbol{\beta}$ in the form of *statistical models*. For example, often $Z$ is described by $N(\xi, u^2(\xi))$, a normal distribution with expectation equal to the measured value $\xi$ of $Z$ and variance equal to the squared standard uncertainty $u^2(\xi)$ of $\xi$, and $\boldsymbol{\beta}$ by $N(\boldsymbol{b}, V_b)$, a multinormal distribution with expectation equal to the estimate $\boldsymbol{b}$ of $\boldsymbol{\beta}$ with covariance matrix $V_b$, although other ("non-normal") choices for the distributions can also be made. The corresponding *measurement model* is given formally by "solving" or "inverting" the forward model to give

$$Y = f(Z, \boldsymbol{\beta}) = f(\boldsymbol{X}), \qquad (2)$$

where $\boldsymbol{X}$, comprising $Z$ and $\boldsymbol{\beta}$, are identified as the input or influence quantities. In practice, the measuring system may comprise $m$ such sensors and the measurement task is to infer a measurand $Y$ given by

$$Y = F(Y_1, \ldots, Y_m, \boldsymbol{\Gamma}) = F(\boldsymbol{X}), \qquad Y_i = f_i(Z_i, \boldsymbol{\beta}_i), i = 1, \ldots, m,$$

where the input quantities $\boldsymbol{X}$ now comprise $Z_1, \ldots, Z_m$, $\boldsymbol{\beta}_1, \ldots, \boldsymbol{\beta}_m$ and possibly additional influence quantities $\boldsymbol{\Gamma}$.

Notice that by building a model directly in terms of the sensor outputs, i.e., of the form

$$Y = G(Z_1, \ldots, Z_m, \boldsymbol{\Theta}), \qquad (3)$$

the need to consider the forward models for the sensors is avoided, and such a model is then *data driven*. Another example of a data driven model in this context is

$$Y = G(\boldsymbol{h}_1(\boldsymbol{Z}_1), \ldots, \boldsymbol{h}_m(\boldsymbol{Z}_m), \boldsymbol{\Theta}), \qquad (4)$$

where $\boldsymbol{h}_i(\boldsymbol{Z}_i)$ denotes a set of features extracted from a time series $\boldsymbol{Z}_i$ of response values provide by the $i^{th}$ sensor, which is the basis of *machine learning* approaches to making inferences based on the sensor responses; see also section 5.

If the sensor is uncalibrated, the measurement task can be to infer $\boldsymbol{\beta}$ from knowledge of $Y$ and $Z$ in the form of statistical models that depend on measured data $(\eta_i, \xi_i)$ for $(Y, Z)$ with associated uncertainty information. The corresponding measurement model is typically derived by optimising a *cost function*. For example, in the case that the $\eta_i$ correspond to the values of standards and are assumed to have negligible uncertainty and the $\xi_i$ are obtained independently with standard uncertainty $u(\xi_i)$, a least-squares cost function is often used, viz.,

$$C(\boldsymbol{\beta}) = \sum_i \left(\frac{\xi_i - \phi(\eta_i, \boldsymbol{\beta})}{u(\xi_i)}\right)^2. \tag{5}$$

The estimate of $\boldsymbol{\beta}$ is defined as the minimizer of that cost function, which in turn defines a measurement model, at least implicitly. Under the assumption that the data errors are distributed normally, the minimizer can be considered to be a maximum likelihood estimate (MLE). Under the additional assumption that the prior distribution for $\boldsymbol{\beta}$ is proportional to a constant, the minimizer can be considered to be a maximum a posteriori (MAP) estimate obtained from a Bayesian analysis, with other forms of prior information influencing the estimate obtained. Different assumptions about the statistical models or probability distributions for the data will imply other forms of cost function.

The mathematical model that describes the relation between the dynamic stimulus $Y(t)$, which acts as input to the sensor, and the corresponding dynamic response $Z(t)$, is called a *dynamic model* and is denoted by

$$Z(t) = \mathcal{H}[Y(t)]. \tag{6}$$

As an example, consider the behaviour of an accelerometer, which can be modelled by the ordinary differential equation

$$\ddot{Z}(t) + 2\delta\omega_0 \dot{Z}(t) + \omega_0^2 Z(t) = \rho Y(t), \tag{7}$$

where $\delta$ denotes the damping coefficient, $\omega_0 = 2\pi f_0$ the resonance frequency, $\rho$ a proportionality constant, and $\dot{Z}(t)$ and $\ddot{Z}(t)$ are, respectively, the first- and second-time derivatives of $Z(t)$. This model is a classic example of a white box model because the parameters can be directly related to physical properties of the sensor. For instance, this model is derived by considering a spring-damper system. The time-dependent $Y(t)$ denotes the acceleration to which the sensor is exposed and $Z(t)$ is the corresponding time-dependent response indicated by the sensor. The sensor behaviour is described equivalently by the ordinary differential equation (ODE) (7), the *transfer function $H(s)$* obtained by applying the Laplace transform to the ODE, and the *impulse response function $h(t)$* obtained by applying the inverse Laplace transform to the transfer function.

A dynamic model is called *linear* when it is linear in its dynamic inputs, i.e., for dynamic quantities $Y_1(t)$ and $Y_2(t)$ and constants $c_1$ and $c_2$ it holds that

$$\mathcal{H}[c_1 Y_1(t) + c_2 Y_2(t)] = c_1 \mathcal{H}[Y_1(t)] + c_2 \mathcal{H}[Y_2(t)]. \tag{8}$$

A dynamic model is called *time-invariant* when it does not change with time, i.e., a time shift in $Y(t)$ results in the same time shift in $Z(t)$:

$$\mathcal{H}[Y(t - t_0)] = Z(t - t_0). \tag{9}$$

The accelerometer is a linear time-invariant (LTI) system because the model for the system is linear and time-invariant.

For an LTI system with impulse response function $h(t)$, the relation between the stimulus $Y(t)$ and the response $Z(t)$ is given by the convolution equation

$$Z(t) = \int_{-\infty}^{\infty} h(t - \tau) Y(\tau) \, d\tau. \tag{10}$$

An equivalent model is the linear state-space model

$$\begin{aligned} \dot{\mathbf{V}}(t) &= C \mathbf{V}(t) + D \mathbf{Y}(t), \\ \mathbf{Z}(t) &= E \mathbf{V}(t) + F \mathbf{Y}(t), \end{aligned} \tag{11}$$

where $C, D, E$ and $F$ are known system matrices, which is often used for systems with multiple inputs and outputs or for networks of dynamic systems. All these models have discrete equivalents, in the form of discrete convolutions and difference equations, to model the relationships between sampled values of the sensor stimulus and sampled values of the sensor response. The determination of model parameters for these kinds of models is well established in signal processing and system theory [Oppenheim 2013] and also known as system identification [Ljung 1998]. It is worth noting that state-space models are also commonly used to model sensor networks of known structure. For instance, in electrical power grids or gas grids state-space system models are used. In the Factory of the Future, though, the knowledge about the sensor network structure is usually not considered in terms of such models based on physical arguments. Instead, data streams from the sensors are taken as inputs of black box or grey box models. This holds true, for instance, also for the two examples considered in section 7.

# 5 Data-driven models (black box models)

## 5.1 Basics

Data-driven modelling does not involve the analysis and mapping of previously known structures, physical interactions, and similar properties of the system. In contrast, the system is only described by the interaction with its environment at the system inputs and outputs. This interaction is captured by systematic observation of the inputs and outputs and transferred into a model using machine learning methods. Since the internal structure of the model is not known such a model is also called a *black box model*. Since no knowledge of the internal structure of the system is required to create the model, this procedure can also be applied in cases where the system structure and the interaction of the components in the system are not or not sufficiently known, e.g., in the case of existing systems whose documentation is incomplete. Data-driven models can also be advantageous if the interactions in the system are difficult to describe or to parameterize, e.g., in the case of strongly nonlinear coupled state variables. An analysis of the system and costly parameter identification becomes unnecessary.

Within data-driven models an implicit reduction takes place in so far as only those variables that are observable as input or output variables in the interaction with the environment are considered explicitly in the system modelling. In contrast to analytical models, the internal state variables are usually not physical quantities but instead the model variables are synthetic quantities and therefore cannot always be interpreted clearly. The above-mentioned aspects of model decomposition and aggregation do not occur in data-driven models since the system to be modelled is not decomposed into sub-models or assembled from these. It is possible, of course, to combine several data-driven models to form a higher-level model — however, this is possible for all models, regardless of the type of model and is not meant by the step of aggregation.

The starting point for the creation of a data-driven model is the knowledge about the input and output quantities in the model. These quantities must be observed in all relevant operating or value ranges, and the dynamics of the quantities, e.g., their transients, must also be observed. The selection of the learning data used to create the model for typical system states and operating ranges implicitly results in a delimitation of the model. The learning of the system behaviour on the basis of typical observations implicitly leads to the fact that this behaviour is stored as a "good state" of the system, so that deviations of the observations from the created model become recognizable. In contrast to analytical models, the activation of specific input functions, e.g., step functions, is usually omitted if such examples of the input variables are not typical for the application under consideration.

An advantage of data-driven modelling is that no explicit physical understanding of the system is required, so that modelling can be carried out even for complex systems or systems whose internal structure is unknown. Since only the inputs and outputs of the system are considered, modelling is also possible while the system is running. By limiting the learning process to observed system states, data-driven models are well suited for detecting anomalies, i.e., significant deviations of the current system state from a desired or normal state.

However, data-driven modelling also has limitations and disadvantages. First, data-driven modelling needs large amounts of data with suitable quality. When observing the system, typical values of inputs and outputs must be available in all relevant forms. This is especially the case for industrial applications, where a process is operated only within certain parameter ranges (leading to stable production and high efficiency). In addition, if annotated data is required to map the relation between input parameters and output, such annotations must be available and reliably correct. If such comprehensive observations are not available, the model may be incomplete, i.e., the reduction is too strong, bias may occur in the data-driven model, or the delimitation is not correct. To ensure that the relevant quantities are observed, and that typical observations and system states are selected during the observation, a minimum of knowledge of the system is also required, so that expert knowledge cannot be dispensed with completely. A disadvantage when checking models for plausibility is the already mentioned difficulty in interpreting the model variables. Data-driven models usually have a comparatively low generalizability as mentioned above, since the limitation of the learning process to an observed operating range corresponds to a delimitation to exactly this range. However, a low generalizability may be desirable: in the case of anomaly detection, the limitation to a certain error-free operating range is intended, so that in this case a targeted limitation of the learning process to a data set of observations assessed as good takes place. Furthermore, data-driven models are usually only slightly abstract. Since the inner system structure is not known or not used during model creation, the formation of classes of similar systems is difficult and can only be achieved phenomenologically by analysing the interactions of the system with its environment.

## 5.2 Statistical modelling and machine learning

Data-driven modelling, which is also often described interchangeably as *machine learning*, is rooted in statistical risk minimization. Typically, the case of classification and regression are considered in classical textbooks [Friedman 2001]. For the scope of this article, we will focus on the case of regression, i.e., the prediction of continuous variables. Generally, we try to estimate the parameters $\boldsymbol{\theta}$ of a family of functions $f(\boldsymbol{x}, \boldsymbol{\theta})$ that produce the desired variable $y$ from measurements $\boldsymbol{x}$. Typically, this search is formalized as a minimization of the risk that is the expected value of a loss function $L(f(\boldsymbol{x}, \boldsymbol{\theta}), y)$, which describes the cost that is created with the current parameter setting $\boldsymbol{\theta}$:

$$\mathrm{argmin}_{\boldsymbol{\theta}} \, E[L(f(\boldsymbol{x}, \boldsymbol{\theta}), y)] = \mathrm{argmin}_{\boldsymbol{\theta}} \int L(f(\boldsymbol{x}, \boldsymbol{\theta}), y) \, \mathrm{d}P(\boldsymbol{x}, y). \tag{12}$$

A major problem with this formulation is that $P(\boldsymbol{x}, y)$ is generally not observable, but only $N$ training instances $(\boldsymbol{x}_n, y_n)$. Therefore, statistical learning focusses on the minimization of the empirical risk

$$\mathrm{argmin}_{\boldsymbol{\theta}} \, E[L(f(\boldsymbol{x}, \boldsymbol{\theta}), y)] = \mathrm{argmin}_{\boldsymbol{\theta}} \frac{1}{N} \sum_n L(f(\boldsymbol{x}_n, \boldsymbol{\theta}), y_n). \tag{13}$$

Bayesian approaches allow the integration of further prior knowledge into the above optimization problem. Using the maximum *a posteriori* formulation, the empirical risk can be formulated as

$$\mathrm{argmin}_{\boldsymbol{\theta}} \, E[L(f(\boldsymbol{x}, \boldsymbol{\theta}), y)] = \mathrm{argmin}_{\boldsymbol{\theta}} \frac{1}{N} \sum_n L(f(\boldsymbol{x}_n, \boldsymbol{\theta}), y_n) + \lambda R(\boldsymbol{\theta}) \tag{14}$$

including a regularisation term $R(\boldsymbol{\theta})$ that allows to embed prior knowledge on $\boldsymbol{\theta}$, e.g., sparsity constraints or other properties that are known about the problem.

Using this framework, a multitude of functions can be fitted. Generally, there is no function, machine learning approach, or learning method that can generally be considered as the "best" as the "no-free-lunch" theorem demonstrates [Duda 2012]. Superiority of a certain method can only be shown with respect to a particular problem domain or data set. For a given data set and problem, solution strategies may vary. When designing machine learning models, it is important to keep the so-called bias-variance trade-off in mind. It tells us that the expected deviation of the model from the training data $E[(y - f(\boldsymbol{x},\boldsymbol{\theta}))^2]$ can be decomposed into a bias term $B[f(\boldsymbol{x},\boldsymbol{\theta})]$ and a variance term $V[f(\boldsymbol{x},\boldsymbol{\theta})]$:

$$E[(y - f(\boldsymbol{x},\boldsymbol{\theta}))^2] = B[f(\boldsymbol{x},\boldsymbol{\theta})]^2 + V[f(\boldsymbol{x},\boldsymbol{\theta})] + \sigma^2, \tag{15}$$

where $\sigma$ is an irreducible error term. The bias is found as

$$B[f(\boldsymbol{x},\boldsymbol{\theta})] = E[f(\boldsymbol{x},\boldsymbol{\theta}) - y]$$

and the variance is determined as

$$V[f(\boldsymbol{x},\boldsymbol{\theta})] = E[(f(\boldsymbol{x},\boldsymbol{\theta}) - E[f(\boldsymbol{x},\boldsymbol{\theta})])^2].$$

Hence, we can conclude that the error on the training set, i.e., the bias, can always be reduced by an increase in variance. This trade-off tells us that an increase in parameters of the training model, corresponding to an increase in model capacity, will be able to reduce the training error. However, this performance increase on the test set comes at the cost of increased variance of the model. A result is typically that the performance on an unseen test set is weak, which is commonly referred to as overfitting. This general result holds for all machine learning models and the only known method to reduce this trade-off is the incorporation of prior knowledge [Duda 2012].

The above observations gave rise to many different machine learning methods that have been developed over the last decades. Important instances include multivariate regression, principal component analysis, decision trees and random forests, ensembling, boosting, as well as support vector machines.

## 5.3 Application of neural network approaches

Recently, artificial neural networks have sparked a "deep revolution", as it could be demonstrated that several previously unsolved tasks could be tackled. Most noteworthy are the examples of speech recognition, image classification, and application of reinforcement learning techniques to board games with high branching factor such as Go [LeCun 2015]. The significant steps taken in these examples can be attributed to availability of large public data sets, open source software, tremendous increases in compute power, as well as a few methodological advances such as differentiable convolutions, and non-saturating activation functions such as the rectified linear unit (ReLU). Today, use of such networks is widespread ranging from classification and regression tasks to inverse problems such as CT reconstruction and applications in simulation of fluids and particles [Maier 2019a].

As an example, Figure 4 demonstrates the information retrieval for multispectral light field cameras [Schambach 2022]. Such cameras are meant to capture spectral information (with about a dozen of channels) and depth information (in the form of disparities) simultaneously [Schambach 2022]. However, reconstructing the spectral and the spatial properties of the scene from the recorded images is a challenging task, since in the images, the spectral and spatial information is implicitly combined. One approach to solve this problem is to use a data-driven modelling to learn the connection between the acquired image data and the desired output data. To this end, training data in the form of multispectral light fields with known ground truth must be available [Schambach 2020]. Here, care must be taken not to introduce undesired bias in the training data, e.g., with respect to the distribution of spectral values and spatial positions. The results depicted in Figure 4 show that it is possible to use such data-driven models to reconstruct the desired spectral and spatial information with high quality.

     A major problem for these deep learning methods is their large number of parameters that makes them inherently black box models. Approaches to attempt interpretations exist, yet they are largely connected to projection of gradients into the input domain and selection or generation of mostly activating input images [Maier 2019a]. Trust and generalization ability are a major concern for such models [Huang 2018].

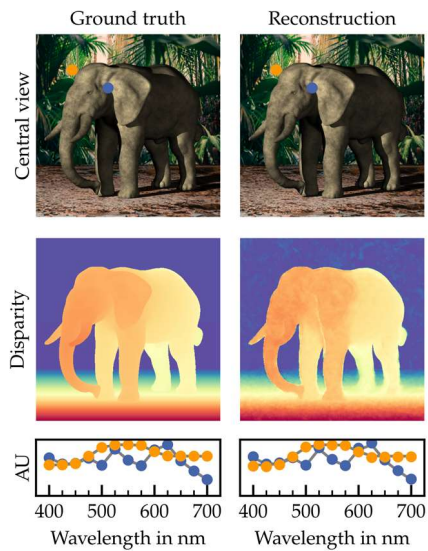

Figure 4: Reconstruction of spectral and spatial information from multispectral light fields using data-driven modelling. RGB representation of the reconstructed spectral information (top), reconstructed spatial information (depicted as disparity, middle), exemplary spectra of two scene points (bottom). The columns show the ground truth (left) and the data-driven reconstruction.
Image source: Maximilian Schambach, KIT.

# 6 Combined modelling approaches (grey box models)

## 6.1 Basics

Analytical and data-driven modelling are not necessarily in competition with each other but are sometimes combined to take advantage of both approaches. One important reason for mixing the approaches in practice is that the choice of approaches often depends on the professional background of the actors. While engineers, e.g., tend to want to understand a system and therefore tend to use analytical models, computer scientists are used to handling data, so they often prefer to use data-driven models.

Even if the physical structure of the system is defined from analytical modelling and the model parameters are initially identified by analysing the system, the quality of the parameter adjustment can usually be improved by a later data-driven optimisation. This approach of "data-driven optimised analytical modelling" can be applied advantageously, such as when the internal state variables of the system are known but are coupled by unknown or difficult-to-identify interactions, e.g., those that are strongly non-linear. With a pure white box model approach the unknown interactions would need to be incorporated by means of probabilistic arguments or as model uncertainties. A combined approach of analytical and data-driven modelling may then reduce the overall uncertainty.

In the case that the modelling of a system is to be essentially data-driven, it can nevertheless be advantageous to first structurally break down the overall system into a few, easily identifiable sub-systems. Depending on the type of sub-systems, this approach of "analytically structured data-driven modelling" can have the advantage that the resulting sub-systems can be more easily described in several, initially separately considered models.

A recent paper found that incorporation of known sub-systems is generally favourable in terms of maximum error bounds [Maier 2019b]. The paper demonstrates that inclusion of prior knowledge in terms of differentiable modules always reduces the maximum error that is produced by solving the learning problem. The authors demonstrate the applicability of their theory in grey box models integrating differentiable projectors into trainable CT reconstruction, measurement of retinal vessels using hybrids between deep learning and classical image processing, as well as for the application of image re-binning. As such the trend of inclusion of differentiable known modules in deep networks can also be related to a solid theoretical basis and demonstrates that indeed classical theory and novel data-driven methods are not in conflict. Instead, solutions combining the best of both worlds are preferable. Deep networks can even be reverse engineered in order to identify relevant processing modules within the deep network as shown in [Fu 2019].

## 6.2 Digital twins in measurement

Digital twins, as a particular instance of grey box models, are one of the most important and up-to-date enablers for automation and especially for autonomous systems. But they are also playing an increasingly important role in measurement of complex processes or objects, e.g., the geometry of multidimensional structures or the combination of multiple measured variables, e.g., in indirect measurements, soft sensors and condition monitoring.

In general, a digital twin is a complete digital representation in the virtual world of an object or process from the real world. This digital representation contains selected features, states and, depending on external influences, behaviour of the object or process. It is therefore a system of aggregated digital models that are controlled by processed external data and information. Within this representation, different models, information, or data can be linked to each other during different life cycles or process phases [Stark 2020]. Connectivity and interoperability with IoT (eco-) systems are therefore essential capabilities of digital twins. Digital twins thus enable an extensive, and importantly for measurement technology and metrology, largely uncertainty-free data exchange. Data to be passed on are theoretically known to be very precise, except for discretization uncertainty.

A digital twin is created as a complex model from a real instance, e.g., a measurement object or measuring system, or from a so-called digital prototype/master. The preparation of the operating, status or process data of the real instance is often referred to as a *digital shadow*. A digital shadow has the task of converting large amounts of heterogeneous data into a form that can be easily transferred to and used in specialized metrological models. The data sets generated in this way meet the following criteria [Brecher 2020]: domain and application specific aggregation; multi-perspectivity; data access rights and private data protection; persistence of data sets, traceability, and uncertainty; semantic enrichment and selection; data cleansing, quality assurance and characterization.

Figure 5 shows the principle operation of a digital twin in measurement and sensor technology and metrology. Figure 6 illustrates the application of digital twins to the quality control of a complete vehicle chassis with respect to geometric dimensions, shape deviations, surface quality, colour tone and possible defects. For this purpose, a virtual chassis is used as a master model, which includes thermal behaviour, material properties, stress behaviour, among other things. Empirical uncertainty values are used to evaluate the quality of the chassis.

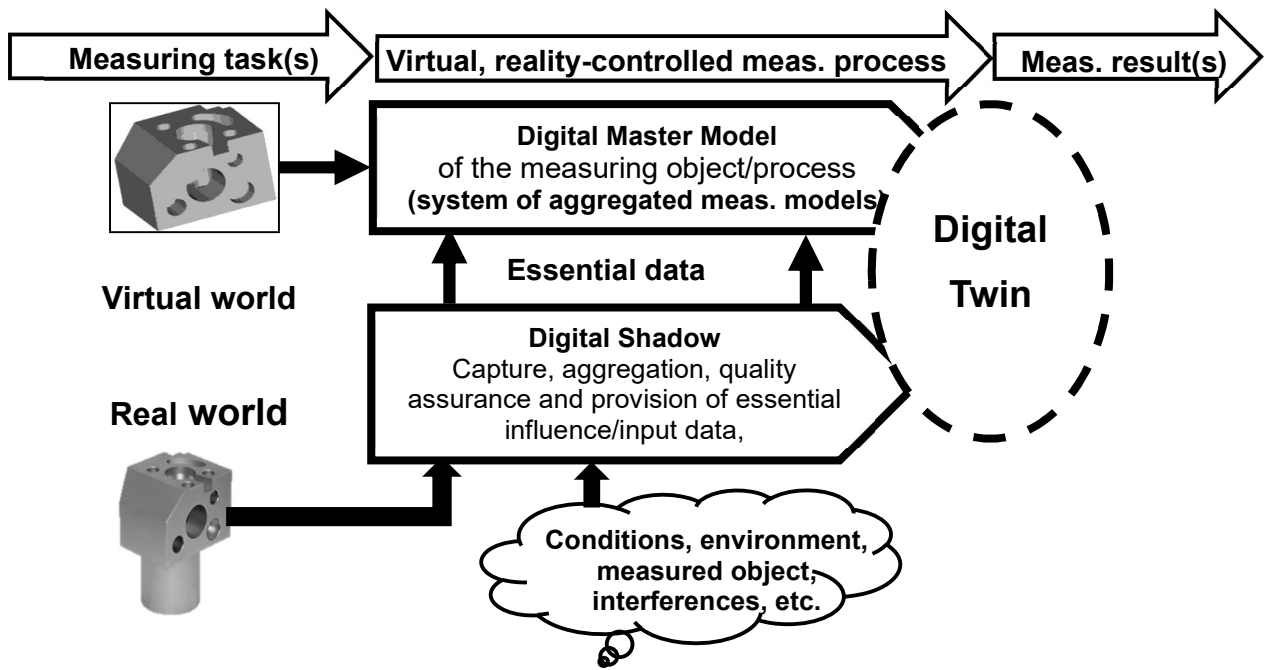

**Figure 5:** Principle operation of a digital twin in measurement and sensor technology.

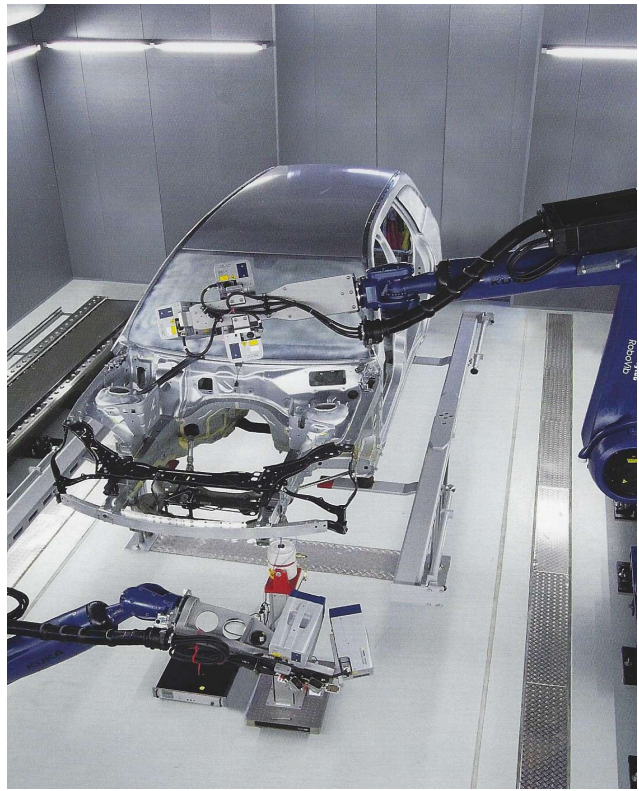

**Figure 6: Application of digital twins to the quality control of a complete vehicle chassis (source: Polytec, Germany).**

# 7 Examples

## 7.1 Modelling for condition monitoring

Condition monitoring is a classification or regression method to derive a statement on the state-of-health or remaining lifetime of an asset based on measurements by a network of sensors. The model for condition monitoring describes the steps from the raw data obtained using the sensor network to the result of the classification or regression. For instance, the model can describe the steps of feature extraction, followed by feature selection and finally the actual classification. The output of the model does not necessarily correspond to a quantity in the metrological sense, i.e., as a property with a magnitude that can be expressed by a number and a reference [VIM 2012]. Instead, the classification simply provides a statement on the class membership.

The correct choice of feature extraction and selection methods depends on the situation, sensor data properties and classification target. At the "Zentrum für Mechatronik und Automatisierungstechnik" (ZeMA) a fully automated toolbox [Schneider 2018] is applied that, based on a cross-validation, determines the model as a combination of feature extraction and selection to provide the best outcome in the training (see Figure 7). The model is thus derived using a complete data driven approach, i.e., it is a black box model, and no physical or other understanding of the underlying processes is assumed. Moreover, for the sensor network providing the input data to the toolbox, no mathematical model of the sensor relations, e.g., as a state-space system model, is required.

A pre-defined set of model building blocks is defined for extracting features from the sensor data, here consisting of the methods listed on the left-hand side of Figure 7. This set of methods as candidates for the model is determined from experience and expertise and could be extended with any other method that extracts parameters from the raw sensor data. A more detailed description of feature extraction is given in [Olszewski 2001].

When the feature extraction methods are picked manually, e.g., based on knowledge about the nature of the sensor data, the model becomes of grey box type. Consider, e.g., one sensor measuring a periodically changing quantity with sinusoidal characteristics. Feature extraction, and even feature selection, could make use of this knowledge by choosing a Fourier transform based feature selection method. Another example is knowledge about redundancy in the sensor network. This knowledge could be used to combine data from the redundant measurements as part of the feature extraction. A further example is the knowledge about synchronisation or timing issues in the network. The correct alignment of the time series to one another is important when the feature extraction relies on valid time stamps. Pre-processing of raw sensor data or selective choice of feature extraction methods that are robust with respect to such effects would mitigate this.

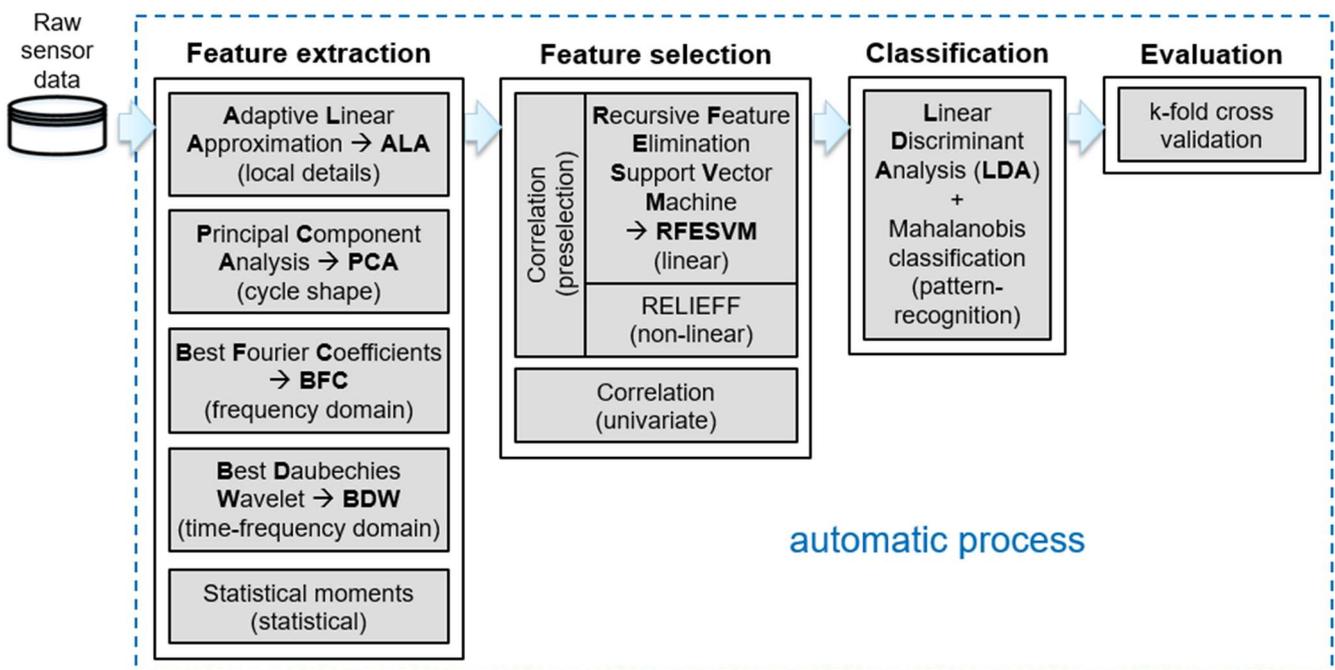

**Figure 7: Schematic of the algorithms for feature extraction, feature selection, classification and validation in the algorithm toolbox [Dorst 2019].**

Usually some knowledge is available about the sensors from which data is obtained. For instance, some sensors may be calibrated or a data sheet from the manufacturer may be available. In these cases, data pre-processing is carried out before features are extracted. For instance, when knowledge of the dynamic behaviour of a sensor in terms of a corresponding white box model (parametric or non-parametric) is available, a deconvolution can be applied to remove dynamic errors. It is worth noting that the whole machine learning toolbox given in Figure 7 does not require such pre-processing. However, transferability, reproducibility and comparability of its results could be improved by using a grey box model approach based on such knowledge.

The third step in the toolbox, a further dimensionality reduction followed by the actual classification, uses the selected features and the training data. The methods used in the toolbox for this step lead to a completely data-driven black box approach. When further knowledge is available, though, other methods can be applied instead. For instance, a Bayesian classification could be employed that makes use of prior knowledge about the underlying characteristics of the problem.

## 7.2 Modelling for quality prediction

Radial forging is widely used in industry to manufacture parts for a broad range of sectors including automotive, aerospace, rail and medical. The Advanced Forming Research Centre (part of the University of Strathclyde, Glasgow, UK) houses a radial forge that uses two pairs of hammers operating at 1,200 strokes/min with a maximum forging force per hammer of 1,500 kN to manufacture parts in a predefined form. The hammers are concave to match part specification, and the rotational speed of the part in the forge is approximately 44 rpm with a programmable range of 40–70 rpm. Both hollow and solid material can be formed with the added benefit of creating internal features on hollow parts using a mandrel. Parts can be formed at a range of temperatures from ambient up to 1,200 °C. Typical materials forged include steel, titanium, and nickel alloys.

The forge includes about 100 sensors recording data at a temporal frequency of 100 Hz for a range of different quantities, both digital, e.g., a power being turned on and off, and analogue, e.g., temperature. They include those dedicated to process output and those dedicated to machine maintenance and are used to provide information about both the *heating phase*, during which the part is heated to a specified nominal temperature for forging, and the *forming phase*, during which the part is forged into the desired form. The hammers are aggregated into two sensor diagnostics (left and right) that record hammer force, position, and speed. Sensors that are of particular interest are those providing information about the temperature and speed of motion of the part as it is forged and those providing information about the forces applied by the hammers. There is appreciable redundancy and correlation in the sensor measurements. An initial assessment indicates that twenty of the sensors contribute no information for the geometrical analysis or energy consumption tasks described below (but they may be useful for other tasks and for diagnosis), and the signals provided by the other sensors are highly correlated. The sensor signals providing data about the nominal and actual values of a quantity are also highly correlated, and in this case the difference between the signals provides more useful information.

One key target process output is energy consumption for the forging of each part, which contributes to manufacturing cost. Experiments show that this output depends strongly on the temperature of the workpiece at the entry to the forge following the heating phase, and the axial feed speed of the chuck head and the radial feed speed of the hammers during the forming phase. Other key process outputs include the geometric dimensions of the parts, which are shown also to be highly correlated and to contain redundant information about the geometry of a workpiece. Furthermore, it is expected that different sections in time of the sensor signals relating to the forging of a part, with these sections delimited by particular start and end times, provide information about different geometric characteristics and dimensions of the part.

Relating the target process outputs to the sensor signals involves stages of dimensionality reduction, feature extraction and selection, and regressing values of the target process outputs collected in a training dataset against the selected features. A mathematical model describing the relationships between the target process outputs and the sensor signals is constructed by combining models for these different stages. For example, the stage of dimensionality reduction can be undertaken using methods of principal component analysis (PCA), independent component analysis (ICA), clustering or autoencoding, all of which are data-driven. Furthermore, extracted features can typically include statistical measures, such as signal maximum, mean or variance, that are not necessarily motivated by expert knowledge [Barandas 2020, Christ 2016]. Finally, regression models such as Random Forests or Support Vector Machines can be used to predict final workpiece dimensions.

Due to the complexity of the forging process, comprising both heating and forming phases, it is difficult to describe the complete process using explicit, physically based mathematical models and so a white box approach is generally not possible. On the other hand, because training datasets, comprising sensor signals and corresponding target process output values for a collection of parts, are generally small, e.g., limited to less than 100 parts by budgetary constraints, a pure black box or data-driven approach to the modelling is often not successful in providing models that generalise well to new test data. In this case, grey box modelling might be a preferred approach in which available data is supplemented by a modest amount of knowledge coming from a physical understanding of the process. For example, certain features, such as the temperature drop between the two phases of forging, extracted from certain sensor signals are expected to provide useful information about target process outputs and can be used to guide the modelling activity.

Data collected from a forging task has been made available [Tachtatzis 2019] and a detailed study of the data is presented in [Luo et al 2021]. The study illustrates the different stages of the data processing pipeline, including (a) pre-processing of the sensor signals to extract information about the heating and forming phases and their alignment for the different parts, (b) exploiting redundancy in the sensor measurements, (c) feature extraction and selection, and (d) prediction of target process outputs based on regression models applied to the selected features.

As an indicative example, consider the target process output of energy consumption for the forging of a part, defined by

$$E = \int_{T_0}^{T_1} p(t)\, dt,$$

where $p(t)$ is the output of the single sensor measuring "power" and $[T_0, T_1]$ is the time interval over which the forging takes place. Figure 8 shows values of energy consumption for the parts forged in three runs: the batch of 81 parts available from [Tachtatzis 2019], and two subsequent batches for which three controlled variables for the forging are varied systematically in order to generate a wide range of process output values for training a predictive model. Using features selected from the set of sensor signals, preliminary results show that well used algorithms, such as Random Forest Regression, provide good prediction results, as shown in Figure 9, with an $R^2$ value of over 0.9.

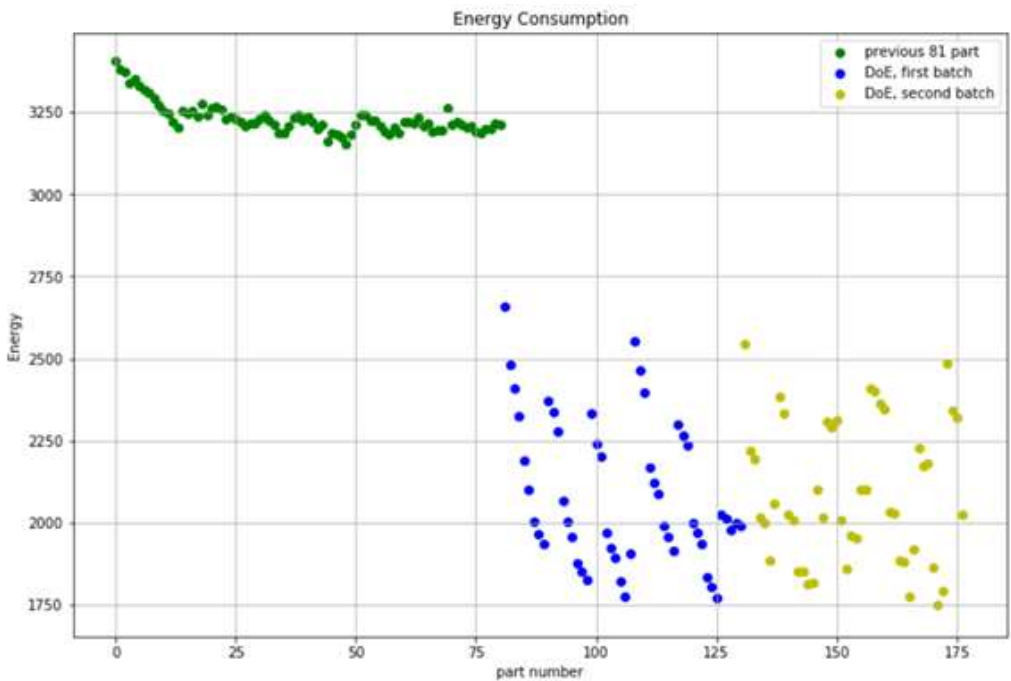

**Figure 8: Values of the target process output of energy consumption for parts forged in three runs.**

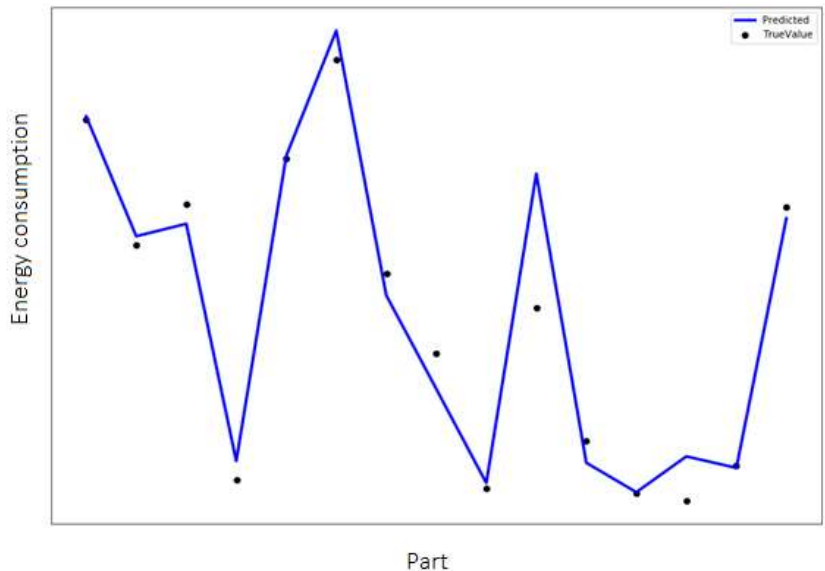

**Figure 9: Prediction of the target process output of energy consumption for out-of-sample data.**

# 8 Conclusion and outlook

Mathematical modelling is a core task in the analysis of measurements. The model represents a relation between observed values and the underlying mechanisms and resulting conclusions derived from the observed data. This relationship applies for the case of simple measurements of a univariate quantity as well as for complex measurements comprising measuring systems of distributed sensors. The theory of mathematical modelling of measurements is well established and a wide range of methods is available in the literature.

In metrology the mathematical model that relates the measured values to (an estimate of) the value of the measurand plays an essential role in the evaluation of uncertainty. This model is the basis for the propagation of uncertainties associated with the influencing quantities to an uncertainty associated with the estimate of the measurand. A recently published supplement [GUM6 2020] to the "Guide to the Expression of Uncertainty in Measurement" is dedicated to the task of modelling, providing an overview of several methods typically applied in metrology.

The Factory of the Future is characterised by networked, distributed measuring systems, large volumes of volatile data, i.e., so called "big data" and data-driven modelling approaches. The combination of these characteristics raises challenges for the well-established modelling approaches typically applied in metrology. For instance, the supplement [GUM6 2020] introduces linear time-invariant system models for the measurement of time-dependent quantities. However, in the Factory of the Future many such sensors are typically combined with other sources of data into a complex network of sensors.

In this contribution we briefly outlined the existing concepts from modelling and discussed their potential use for scenarios for data analysis in the Factory of the Future. White box, black box and grey box models are all well known in system identification and other fields. However, their use in metrology as the basis for the propagation of uncertainty in sensor networks has not been addressed until recently. This paper aims to provide a starting point for such a development by introducing basic concepts, highlighting potential future research and outlining concepts for the treatment of uncertainty. The use of white and grey box models is quite well known in metrology with, e.g., the recent supplement to the GUM focusing on these model types. The extension and further development of the underlying principles towards the treatment of black box, data-driven models will build upon this knowledge. Therefore, this paper provides the basic principles for white and grey box models and introduces their distinction to black box models.

The two examples provide first insights into the components of mathematical modelling in sensor networks for addressing the evaluation of uncertainties. Future research will combine mathematical modelling approaches in sensor networks with the propagation of uncertainties in accordance with the GUM and its supporting documents. Due to the complex nature of practical sensor networks and the volatility of the measured data, these approaches will combine several of the modelling principles outlined in this paper.

# Acknowledgement

Part of this work has been developed within the Joint Research project 17IND12 "Met4FoF" of the European Metrology Programme for Innovation and Research (EMPIR). The EMPIR is jointly funded by the EMPIR participating countries within EURAMET and the European Union.